\documentclass[useAMS,usenatbib]{mnras}
\usepackage{hyperref}
\usepackage{multicol}
\usepackage{longtable}
\usepackage{supertabular}
\usepackage{epsfig}
\usepackage{lineno}
\usepackage{lscape}
\usepackage{graphicx}
\usepackage{amssymb}
\usepackage{rotating}
\usepackage{natbib}
\usepackage{graphics}
\usepackage{amsmath}

\def\lta{\mathrel{\spose{\lower 3pt\hbox{$\mathchar"218$}} \raise
2.0pt\hbox{$\mathchar"13C$}}} \def\gta{\mathrel{\spose{\lower
3pt\hbox{$\mathchar"218$}} \raise 2.0pt\hbox{$\mathchar"13E$}}}

\title[]{Broad-band spectral study of X-ray transient MAXI J1820+070 using {\it Swift}/XRT and {\it NuSTAR}}
\author[Bharali et al.]{Priya Bharali$^{1,\,2}$\,\thanks{E-mail:priya\_bharali@gauhati.ac.in}, Jaiverdhan Chauhan$^{3}$ and Kalyanee Boruah$^{1}$ \\
$^{1}$ Department of Physics, Gauhati University, Guwahati 781014, India.\\
$^{2}$ Mahatma Gandhi Government Arts College, Mahe, Puducherry 673311, India\\
$^{3}$ International Centre for Radio Astronomy Research, Curtin University, GPO Box U1987, Perth, WA 6845, Australia.\\
}
\begin{document}
\begin{NoHyper}

\date{}

\pagerange{\pageref{firstpage}--\pageref{lastpage}} \pubyear{2018}

\maketitle

\label{firstpage}

\begin{abstract}
We report on a \textit{NuSTAR} and \textit{Swift}/XRT observation of the newly discovered X-ray transient MAXI J1820+070. \textit{Swift}/XRT and \textit{NuSTAR} have concurrently observed the newly detected source on 14 March 2018. We have simultaneously fitted the broad-band spectra obtained from \textit{Swift}/XRT and \textit{NuSTAR}. The observed joint spectra in the energy range 0.6--78.0 keV are well modeled with a weak disk black-body emission, dominant thermal Comptonization and relativistic reflection fraction. We have detected a fluorescent Iron-K$\alpha$ line relativistically broadened, and a Compton hump at $\sim$ 30 keV. We constrain the inner disk radius as well as the disk inclination angle and their values are found to be 4.1$^{+0.8}_{-0.6}$ R$_{ISCO}$ (where R$_{ISCO}\equiv$ radius of the innermost stable circular orbit) or 5.1$^{+1.0}_{-0.7}$ r$_{g}$ (where r$_{g}\equiv$ gravitational radius) and 29.8$^{+3.0}_{-2.7}$ $^\circ$ respectively. The best fit broad-band spectra suggest that the source was in the hard state and evolving. The source emission is best described by weak thermal emission along with strong thermal Comptonization from a relatively cold, optically thick, geometrically thin and ionized accretion disk. X-ray spectral modeling helps us to understand the accretion and ejection properties in the vicinity of the compact object.
\end{abstract}


\section{Introduction \label{sec:Intro}}
\hspace{0.8cm}A new bright uncatalogued X-ray transient MAXI J1820+070 was discovered by the Monitor of All-sky X-ray Image\footnote{\url{http://maxi.riken.jp/top/index.html}} \citep[{\it MAXI};][]{Matsuoka2009} on 11 March 2018 \citep{Kawamuro2018}, ASASSN-18ey was detected as the optical counterpart of MAXI J1820+070 \citep{Denisenko2018}. Subsequent observations in X-ray and optical indicate that the compact object is a strong black hole candidate \citep{Baglio2018, Uttley2018}. Initial monitoring reported that the source was in the hard state \citep{Uttley2018, Del2018, Bozzo2018}. This transient was found to exhibit a chaotic variability on the timescale of fewer than 5 seconds. Low-frequency quasi-periodic oscillations were detected around 40 and 58 mHz \citep{Mereminskiy2018, Yu2018}. In the optical band, the source showed fast flaring of the order of msec to sec \citep{Littlefield2018, Sako2018, Gandhi2018}. The outburst was also detected in radio frequency \citep{Bright2018, Trushkin2018, Tetarenko2018}. \citet{Kara2019} performed spectro-temporal study of MAXI J1820+070 and detected reverberation timelags between 0.1--1.0 keV and 1.0--10.0 keV energy bands. The authors have observed corona height $<5$ r$_{g}$ and inner disk radius $<2$ r$_{g}$. \citet{Kara2019} have also explained that with the evolution of the outburst the corona becomes compact and shifts close to the compact object. During the time of this study, MAXI J1820+070 was still in outburst and evolving.

\hspace{0.8cm}During a typical outburst, black hole X-ray binaries (BHXRBs) undergo different state transitions, from the low/hard (LH) to the high/soft (HS) state through relatively short-lived intermediate states (HIMS and SIMS) \citep{Bel00, Remillard2006}. The changes in the accretion flow geometry in the vicinity of the black hole are believed to be associated with these state transitions. The inner disk is supposed to extend to the innermost stable circular orbit (ISCO) in the soft state. Estimations of the black hole spin are possible via X-ray spectroscopy, as the location of the ISCO is defined by the black hole angular momentum. These estimations can be done either by modeling the thermal disk \citep{Zhang1997} or the disk reflection component \citep{Fabian1989}. Furthermore, information about the nature of the illuminating source commonly referred to as ``corona" can be obtained from the relativistic reflection spectrum \citep{Miller2015, Walton2017}. These relativistic reflection spectra of X-ray binaries are the signatures of interaction between radiation emitted near the compact object and the nearby cold gas. The effects of this reflection include iron $K_\alpha$ line emission in the range 6-8 keV \citep{Miller2007}. The $K_\alpha$ line is observed from nearly all accreting compact sources \citep[e.g.][]{Got95, Bharali2019, Kara2019}. Asymmetry in the Fe $K_\alpha$ line profile has been observed in many sources with the advancement in X-ray spectroscopy. The comprehensive study of these characteristics can be used to understand the Doppler and gravitational redshifts, thereby giving crucial information about the environment and kinematics of the cold gas. The flattening of the spectrum above 10 keV is usually called the high energy bump or the Compton hump, arises due to the photoelectric absorption and the Compton up-scattering of the thermal seed photons in the accretion disk by the hot electrons of the corona \citep{Fabian2000, Miller2007}.

\hspace{0.8cm}In this present investigation, we provide a broad-band spectral view of MAXI J1820+070, observed during its recent outburst from {\it Swift}/X-ray Telescope (XRT) and \textit{NuSTAR} observatories. With high sensitivity, broad bandpass and triggered readout free from pile-up distortion, Nuclear Spectroscopic Telescope Array \citep[\textit{NuSTAR},][]{Harrison2013} is ideal for studying reflection in Galactic binaries. \textit{NuSTAR} provides pile-up free performance (up to 100 mCrab), high energy resolution ($\sim400$ eV in the range 0.1--10 keV) and excellent calibration. A rare opportunity to study the relativistically skewed Iron line profiles along with a constrained inner disk radius with great precision is provided by such an instrument. The paper is organized in the following manner. Section $\S$1 and $\S$2 present the introduction and the details of observations \& data reduction methodologies, respectively. Section $\S$3 refers to the analysis \& results of the aforementioned outburst. The last section $\S$4 presents the discussions and conclusions of the major obtained results.

\section{Observation and Data Reduction \label{sec:ObsRedcut}}
                                  
\begin{table*}
 \centering
 \caption{Monitoring details for the {\it Swift}/XRT and {\it NuSTAR} on MAXI ~J1820+070.} 
\begin{center}
\scalebox{1.0}{%
\begin{tabular}{ |l|c|c|c|c|c|c|c| }
 \hline
\hline
Instrument  & Observation  & Observation & Observation & MJD  & Effective & Count rate & Observation \\
 Name &  ID  & Start date & Start time &  & exposure & (cts s$^{-1}$) & mode \\
  &  & (DD-MM-YYYY) & (HH:MM:SS) & & time (ks) &  &  \\
\hline
NuSTAR/FPMA, FPMB & 90401309002 & 14-03-2018 & 20:26:09 & 58191.85  & $\sim11.77$ & $527\pm22$ & SCIENCE \\
\hline
SWIFT/XRT & 00010627001 & 14-03-2018 & 20:56:32 &  58191.87 & $\sim1.04$ & $167\pm12$ & WT \\
\hline
\hline
 \end{tabular}}
 \end{center}
\label{tab:tab1}
\end{table*}

\subsection{{\it NuSTAR} FPMA and FPMB}
\hspace{0.8cm}{\it NuSTAR} has observed MAXI J1820+070 on March 14, 2018 (Obs. ID:90401309002). Two focal plane module telescopes (FPMA and FPMB) were used to acquire the {\it NuSTAR} data. The net effective on-source exposure after detector dead time correction is 11.77 ks. The \textit{NuSTAR} data are processed using the NuSTARDAS software (v1.7.1) included in the {\tt HEASOFT} v6.23 along with the latest calibration files CALDB v20190513. The {\tt nupipeline} task (version 0.4.6 released date : 2016-10-06) is used for filtering event files. The spectra were extracted from a circular region of the radius 100 arcsec centered on the source location. The location of the circular region was fixed after inspecting the source position in the observed image for both the detectors individually. In order to avoid the contamination by source photons, we have chosen the background region (in the detector plane) furthest from the source location.
 
The {\tt NUPRODUCTS} task is used to generate science products such as light curves, energy spectra, response matrix files (rmfs) and auxiliary response files (arfs) for both telescopes FPMA and FPMB. The signal-to-noise ratio is increased by merging the light curves from both telescopes. Further, for minimising the systematic effects, energy spectra from both the detectors were modeled simultaneously. The \textit{NuSTAR} spectra are modeled using XSPEC v12.10.0 with $\chi^2$ statistics.

\begin{figure}
\centering 
\includegraphics[scale=0.34,angle=-90]{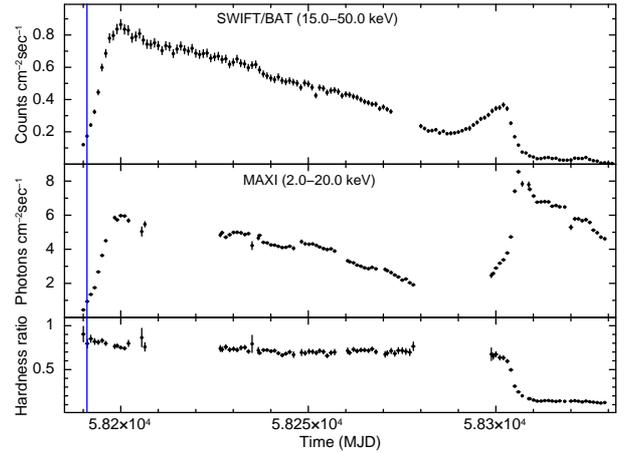}
\caption{{\bf Top panel: } One day averaged {\it Swift}/BAT light curve in the energy range 15.0--50.0 keV. \textbf{Middle panel:} One day averaged light-curve  from {\it MAXI} (2.0$-$20.0 keV) along with hardness ratio \textbf{(Bottom panel)} to display the long-term variations in MAXI J1820+070. The hardness ratio is defined as the ratio of count rates in 4.0--10.0 keV to 2.0--4.0 keV for {\it MAXI} data.}
\label{fig:fig1}
\end{figure}

\subsection{{\it Swift}/XRT}
\hspace{0.8cm}MAXI J1820+070 was also observed independently by {\it Swift}/X-ray Telescope (XRT). This transient was observed by {\it Swift} using normal automated burst response in Windowed Timing (WT) mode, details of which are given in Table \ref{tab:tab1}.

\begin{figure}
\centering 
\includegraphics[scale=0.34,angle=-90]{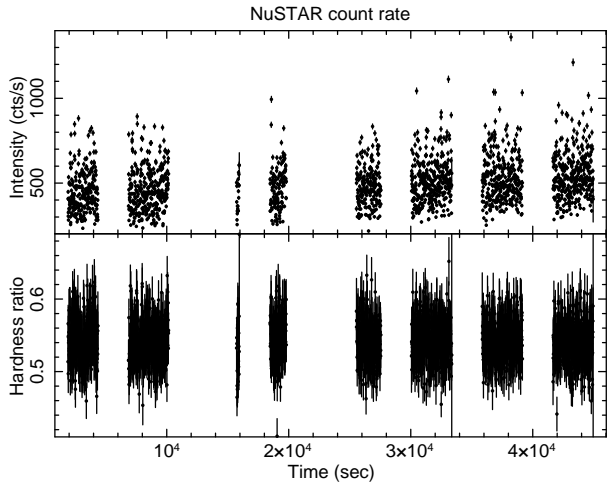}
\caption{The top panel shows 10 sec binned light curve for NuSTAR and the bottom panel shows the variation of hardness ratio. The hardness is defined as the ratio of count rate in the energy bands 10.0--78.0 keV and 3.0--10.0 keV.}
\label{fig:fig2}
\end{figure}

\hspace{0.8cm}Filtering and screening of {\it Swift}/XRT data was done using standard procedures. The {\tt xrtpipeline} version 0.13.4, release date 2017-03-15, was used to reduce the {\it Swift}/XRT data. The background-subtracted average count rate during the observation was found to be $\sim167$ counts sec$^{-1}$. According to \citet{Romano2006}, the specified photon pile-up limit of the WT mode data is $\sim100$ counts sec$^{-1}$. Therefore, following the prescriptions of \citet{Romano2006}, the possibility of the pile-up in the data is tested by looking at the spectral distortion. Before extracting any scientific product, correction for the pile-up was done by removing the appropriate bright portion from the center of the source. In this way, the most suitable pile-up free source region used for our analysis is represented as a circular annulus of inner radius 3 arcsec and outer radius 60 arcsec. A similar circular annulus, except centered at 5 arcmins away from the center along the image strip, was chosen to be the background region. The spectrum and lightcurve corresponding the source and background regions were extracted using ftool {\tt XSELECT} (V2.4e). Afterward, the ARF files were generated using the source spectra and exposure map with the help of the {\tt XRTMKARF} tool. For further spectral analysis, the resulting files are then used along with proper RMF files from recently updated {\tt CALDB} released on 2017-11-13.

\section{Analysis and Results \label{sec:AnaRes}}
\hspace{0.8cm}In the top panel of Figure \ref{fig:fig1}, we have shown one-day averaged light curve of {\it Swift}/BAT (15.0--50.0 keV). A one-day averaged light curve of {\it MAXI} (2.0--20.0 keV) is plotted in the middle panel. The hardness ratio (HR) for {\it MAXI} data is presented in the bottom panel, where HR is defined as the ratio of the source count rates in the energy range 4.0--10.0 keV and 2.0--4.0 keV \citep{Morihana2013}. The {\it NuSTAR} observation is displayed with a blue vertical line. The observation was taken on the third day after the detection of the outburst. In the top panel of Figure \ref{fig:fig2}, we plotted the 10 sec binned light curve for {\it NuSTAR}. We have added the events from FPMA and FPMB to increase the signal-to-noise ratio. In the bottom panel the variation of hardness ratio for {\it NuSTAR} data is shown. The hardness ratio for {\it NuSTAR} data is defined as the ratio of count rate in the energy bands 10.0--78.0 keV and 3.0--10.0 keV \citep{Tomsick2014}.

\begin{figure}
\centering 
\includegraphics[scale=0.34,angle=-90]{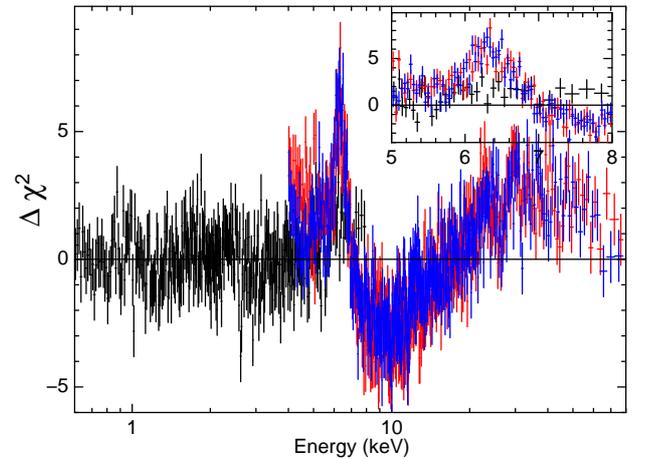}
\caption{The plot shows the observed residuals from the fit using model {\tt TBabs$\times$(diskbb + nthComp)}. A broad Iron line, a dip close to 10 keV and a Compton hump peaking at $\sim30$ keV is clearly observed. In the inset, we have presented iron line region in the energy band 5--8 keV}
\label{fig:fig3}
\end{figure}

\hspace{0.8cm}To understand the spectral characteristics of MAXI J1820+070, the simultaneous fitting of the broadband energy spectra from {\it NuSTAR}/FPMA, FPMB (4.0$-$78.0 keV) and {\it Swift}/XRT (0.6$-$8.0 keV) are performed using tool {\tt XSPEC} version: 12.9.0 \citep{Arn96}. In the case of {\it Swift}/XRT, we have ignored channels below 0.6 keV and above 8.0 keV because of low signal-to-noise ratio (SNR). For {\it NuSTAR}/FPMA and FPMB, channels below 4.0 keV are ignored due to some distortions in the spectra, while channels above 78.0 keV are ignored as there is no response above this energy. All the observed spectra were rebinned so that there are at least 30 counts per energy bin for the spectral fitting.

\hspace{0.8cm} Initially, to precisely explain the observed broadband energy spectra, we fit the data with a model, which is the combination of multicolored disk black-body \citep[{\tt diskbb:}][]{Mitsuda1984, Makishima1986} and thermal Comptonization \citep[{\tt nthComp:}][]{Zdziarski1996, Zycki2009}, {\tt TBabs$\times$(diskbb + nthComp)} in {\tt XSPEC} notations. A typical reflection feature is evident in the broad band energy spectra. The observed residuals from the fit are shown in Figure. \ref{fig:fig3} in which a broad iron line, a dip close to 10 keV and a Compton hump at $\sim30$ keV are clearly visible. We use {\tt TBabs} \citep{Wilms2000} for interstellar absorption due to neutral hydrogen. The abundance input for the {\tt TBabs} model is set to {\it aspl} \citep{Aspl2009}, the most contemporary abundance model inbuilt in {\tt XSPEC}. We use the {\tt CONSTANT} model in {\tt XSPEC} to access the uncertainties associated with the cross-instrument calibrations. We freeze {\tt CONSTANT} at 1 for {\it Swift}/XRT and let it free for {\it NuSTAR}/FPMA and {\it NuSTAR}/FPMB. The value of the {\tt CONSTANT} factor is found to be $1.01\pm0.01$ ($\sim1$\%) and $1.03\pm0.01$ ($\sim3$\%) for {\it NuSTAR}/FPMA and FPMB respectively, which is within the accepted limit of $\leqslant20$\% \citep{Madsen2015, Marcotulli2017}.

\hspace{0.8cm}To account for the observed reflection characteristics, we added a broadband, self-consistent, relativistic reflection model {\tt relxilllpCp} from the relxill \citep[{\tt relxill v1.2.0 :}][]{Dauser2014, Garcia2014} model family. Hence, our best fit model becomes {\tt TBabs$\times$(diskbb+relxilllpCp)}. In the {\tt relxilllpCp} model, it is presumed that the corona is the illuminating source and it is in the lamp-post geometry. In case of the lamp-post geometry, the corona is inferred as a point source located at a height h, above the accretion disk, on the spin axis of the black hole \citep{Miniutti2004}. The thermal Comptonization model, {\tt nthComp} is utilized as the input continuum in the {\tt relxilllpCp} model. In the {\tt relxilllpCp} model, through ray--tracing calculations, the amount of reflection fraction (Refl$_{frac}$) can be self-consistently estimated using the values of the black hole spin parameter (a$_{*}$), the lamp-post height (h) and the inner accretion disk radius (R$_{in}$). The determination of the reflection fraction by the model itself helps in optimizing the parameter space and finally helps in constraining the geometry of the X-ray binary \citep{Dauser2014}.

The introduction of {\tt relxilllpCp} model improves the goodness of fit and the $\chi^2/dof$ is found to be 4134.3/3748 (1.10). The fitted spectra along with the best fit model ({\tt TBabs$\times$(diskbb + relxilllpCp)}) is presented in the left-hand panel of Figure \ref{fig:fig4} and its right-hand panel shows the contribution from various spectral components. The disk temperature (T$_{in}$) is observed to be 0.13$^{+0.01}_{-0.01}$ keV. The observed T$_{in}$ is typical for the hard spectral state \citep[e.g.][]{Wilkinson2009, Petrucci2014, Wang-Ji2018} and close to the value reported by \citet{Uttley2018} using {\it NICER} observations. We have fixed the outer disk radius at 400 r$_{g}$ (where gravitational radius : r$_{g}$ $\equiv$ GM/c$^{2}$). We have concurrently fitted for the spin parameter (a$_{*}$) and the inner disk radius (R$_{in}$) but both the quantities are degenerate in nature. It is very typical to constrain the value of a$_{*}$ and R$_{in}$ simultaneously. The actual value of R$_{in}$ is governed by both a$_{*}$ and disk inclination angle. While fitting the energy spectra, it was observed that a$_{*}$ always reached the hard upper limit of 0.998. Hence, we freeze the value of a$_{*}$ at 0.998 and kept R$_{in}$ free to vary.

Additionally, to constrain the true nature of the system i.e., whether MAXI J1820+070 is favoring truncated disk or low spin, we allowed the spin parameter (a$_{*}$) to vary and we fixed R$_{in}$ at the ISCO. We found that a$_{*}$ continued to peg at the maximum value and when we calculate the 2$\sigma$ uncertainty on a$_{*}$, it gives a lower bound of 0.68. When a$_{*}$ is frozen at 0.998 (maximum value), the $\chi^{2}$/dof is found to be 4134.3/3748 (1.10). However, when we freeze R$_{in}$ at ISCO the $\chi^{2}$/dof is observed to be 4997.5/3748 (1.33). Therefore, this indicates that MAXI J1820+070 favors a truncated disk over low spin. For a black hole of a$_{*}$ $\approx$ 0.998, R$_{ISCO}$ $\approx$ 1.24r$_{g}$ \citep{Bardeen1972}.

The spectral modelling indicated the value of high-energy cutoff (kT$_{e}$) to be 118$^{+34}_{-27}$ keV. It is discovered that the distant reflection component is ionised in nature as Log $\xi$ (\ Log[\ erg cm s$^{-1}$]\ )\ = 3.05$^{+0.01}_{-0.01}$. The absorption column density $n_{\rm H}$ is found to be 0.16$^{+0.03}_{-0.02}$ ( $\times$ 10$^{22}$ cm$^{-2}$), which is close to the galactic absorption value in the line of sight as estimated by online tools by LAB Survey\footnote{\url{https://www.astro.uni-bonn.de/hisurvey/profile/index.php}} \citep{Kalberla2005} ($n_{\rm H} = 0.14 \pm 0.01 \times 10^{22} cm^{-2}$). The source is in the hard state as inferred from the photon-index ($\Gamma$), which is 1.51$^{+0.01}_{-0.01}$ \citep{Remillard2006}. The iron abundance is higher by a factor of almost four as compared to Fe/solar abundance. The Fe/solar abundance is observed to be 4.9$^{+0.4}_{-0.3}$. The lamp-post height (h) is found to be 3.9$^{+0.4}_{-0.3}$ r$_{g}$. \citet{Kara2019} reported the corona height $<5$ r$_{g}$ using {\it NICER} data of MAXI J1820+070, which is close to the lamp-post height observed in this study. The various parameters obtained after fitting are summarised in table \ref{tab:tab2} with 2$\sigma$ error.

The model also provides the inner disk radius (R$_{in}$) and the inclination angle ({\it i}). To precisely constrain the inner disc radius (R$_{in}$) from our best-fitting  model \texttt{($TBabs\times(diskbb+relxilllpCp)$)}, we evaluated $\Delta\chi^{2}$ utilising the {\tt steppar} command in {\tt XSPEC}. The resultant $\Delta\chi^{2}$ with varying R$_{in}$ as the free parameter between 1 R$_{ISCO}$ and 9 R$_{ISCO}$ are shown in the left-hand panel of Figure \ref{fig:fig5}. The two horizontal lines represent the 2$\sigma$ and 3$\sigma$ significance levels. The inner disc radius is found to be 4.1$^{+0.8}_{-0.6}$ (R$_{ISCO}$) or 5.1$^{+1.0}_{-0.7}$ r$_{g}$ (R$_{ISCO}\approx1.24$ r$_{g}$ for a$^{*}\approx0.998$) with 3$\sigma$ significance. Similarly, we have also constrained the disk inclination angle ({\it i}) as shown in the middle panel of Figure \ref{fig:fig5}. The value of {\it i} is found to be 29.8$^{+3.0}_{-2.7}$ degrees, within 3$\sigma$ bounds. We have also explored the correlation between R$_{in}$ and {\it i} using the current best-fit model. In the right-hand panel of Figure \ref{fig:fig5} confidence contours are shown, where 3$\sigma$, 2$\sigma$ and 1$\sigma$ confidence limits are highlighted in blue, green and red colour, respectively.

\begin{figure*}
\centering 
\includegraphics[scale=0.33,angle=-90]{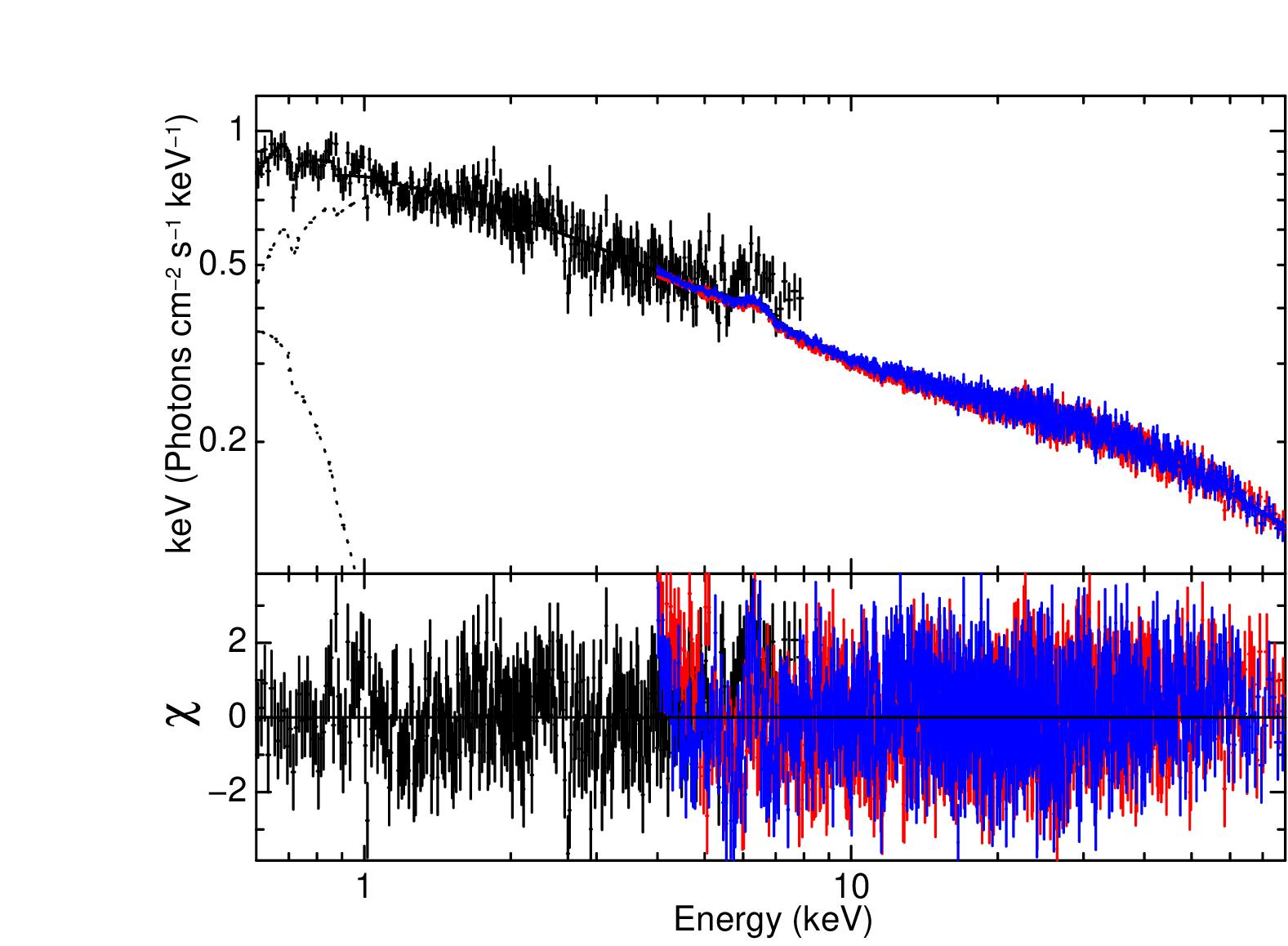}
\includegraphics[scale=0.33,angle=-90]{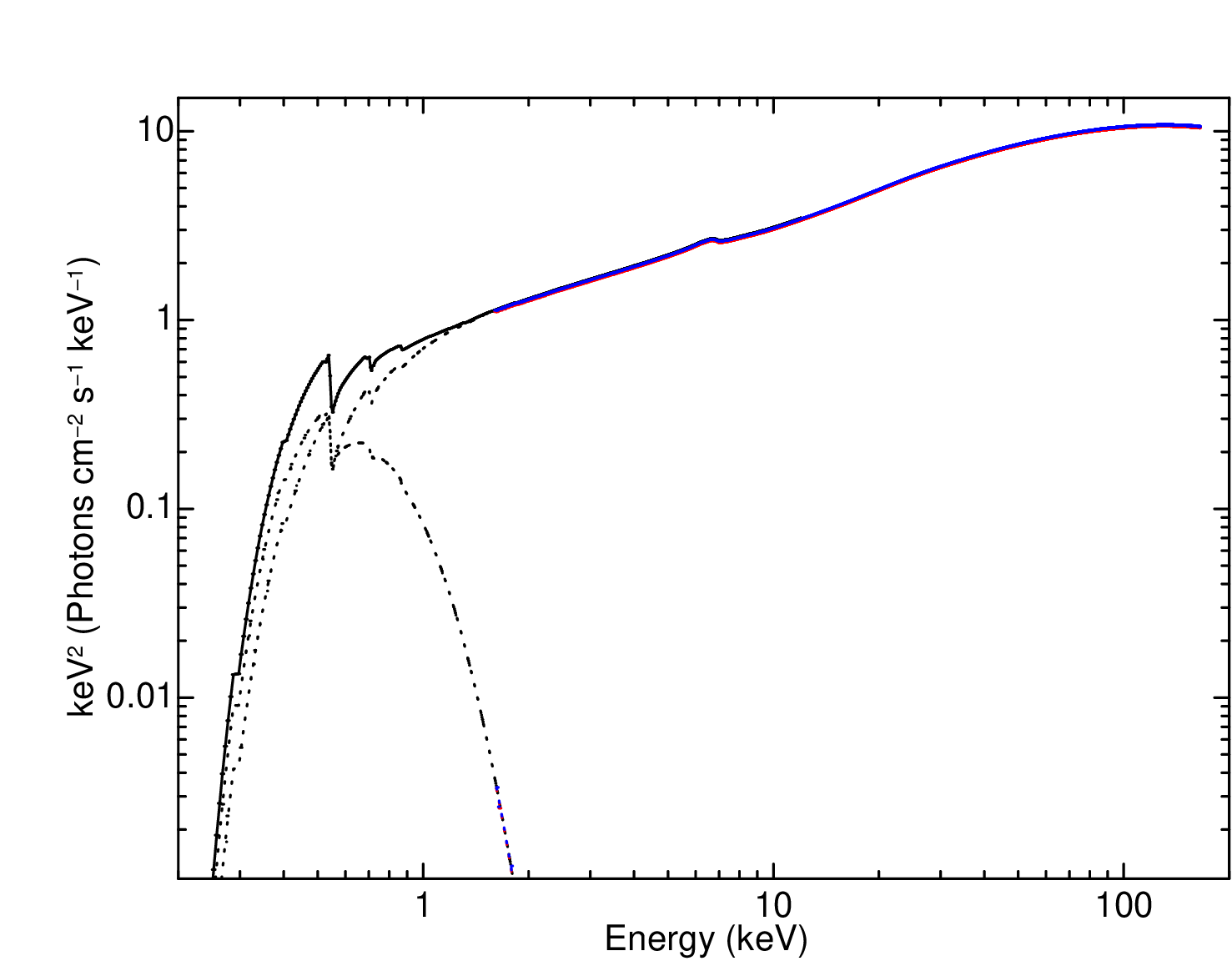}
\caption{The \textbf{left panel} shows the fitted spectra along with the best fit model \ ({\tt TBabs$\times$(diskbb + relxill(lp)Cp)}). The \textbf{right panel} shows the contribution from various spectral components.}
\label{fig:fig4}
\end{figure*}

\begin{table}
\centering
 \caption{Model parameters obtained from the simultaneous {\it Swift}/XRT and {\it NuSTAR} energy spectra fitting. The model {\tt TBabs$\times$(diskbb + relxill(lp)Cp)} gives the best fit with $\chi^2$/dof = 4134.3/3748 (1.10), quoted errors are 2$\sigma$ in nature. Figure. \ref{fig:fig4} presents the best fitted energy spectra along with model constituents and residuals.}
\begin{center}
\scalebox{0.80}{
\begin{tabular}{ |l|l|l| }
\hline
Component & Parameter & Values \\
 &  &  \\ 
\hline
TBabs & $n_{\rm H}$ ($\times 10^{22} cm^{-2}$) & 0.16$^{+0.03}_{-0.02}$ \\
\hline
diskbb & T$_{in}$ (keV) & 0.13$^{+0.01}_{-0.01}$ \\
\hline
relxill(lp)Cp & h (GM/c$^{2}$) & 3.9$^{+0.4}_{-0.3}$ \\
  & a$_{*}$ (cJ/GM$^{2}$) & 0.998 (f) \\
  & i (degrees) & 29.8$^{+1.5}_{-0.8}$ \\
  & R$^{\dagger}_{in}$ (R$_{ISCO}$) & 4.1$^{+0.4}_{-0.3}$ \\
  & $\Gamma$ & 1.51$^{+0.01}_{-0.01}$ \\
  & log $\xi$ (\ log[\ erg cm s$^{-1}$]\ )\ & 3.05$^{+0.01}_{-0.01}$ \\
  & Afe (solar) & 4.9$^{+0.4}_{-0.3}$ \\
  & kTe (keV) & 118$^{+34}_{-27}$ \\
  & Norm. & 0.21$^{+0.02}_{-0.02}$ \\
\hline
* & F$_{Total}$ ($\times$ 10$^{-8}$ ergs cm$^{-2}$ s$^{-1}$) & 2.53$^{+0.02}_{-0.02}$ \\
 & F$_{diskbb}$ ($\times$ 10$^{-8}$ ergs cm$^{-2}$ s$^{-1}$) & 0.049$^{+0.004}_{-0.005}$ \\
 & F$_{relxilllpCp}$ ($\times$ 10$^{-8}$ ergs cm$^{-2}$ s$^{-1}$) & 2.48$^{+0.01}_{-0.01}$ \\
\hline
 & $\chi^{2}$/dof & 4134.3/3748 (1.10) \\
\hline
\end{tabular}}\\
{\bf $\dagger$} : Inner disk radius\\
{\bf f} : Frozen parameters\\
{\bf *} : All the flux values are unabsorbed and calculated in the energy range 0.6--78.0 keV
\end{center}
\label{tab:tab2}
\end{table}

\section{Discussions and Conclusions \label{sec:DiscConc}}
\hspace{0.8cm}In this study, we have performed broad-band spectral analysis of the newly discovered black hole candidate X-ray binary MAXI J1820+070 \citep{Shidatsu2018, Tucker2018} using {\it Swift}/XRT and {\it NuSTAR} simultaneous observations. The detailed spectral analysis revealed that the source was in the hard state during the observation. We observed some hints of soft excess around 1 keV. We have also detected prominent broad Iron line and Compton hump at $\sim30$ keV. The resultant spectrum is well modeled with a multi-color disk black-body, thermal Comptonization, and the self-consistent, relativistic reflection model {\tt relxilllpCp}. We tried to constrain the inner disk radius (R$_{in}$) as well as the disk inclination angle, and their values within 3$\sigma$ limits are found to be 4.1$^{+0.8}_{-0.6}$ R$_{ISCO}$ or 5.1$^{+1.0}_{-0.7}$ r$_{g}$ and 29.8$^{+3.0}_{-2.7}$ degrees, respectively. The disk inclination angle and the inner disk radius were found to be well constrained within $3\sigma$ significance. As the two parameters are degenerate in nature, therefore, variations in one parameter could affect the value of others.

\begin{figure*}
\centering 
\includegraphics[scale=0.22,angle=-90]{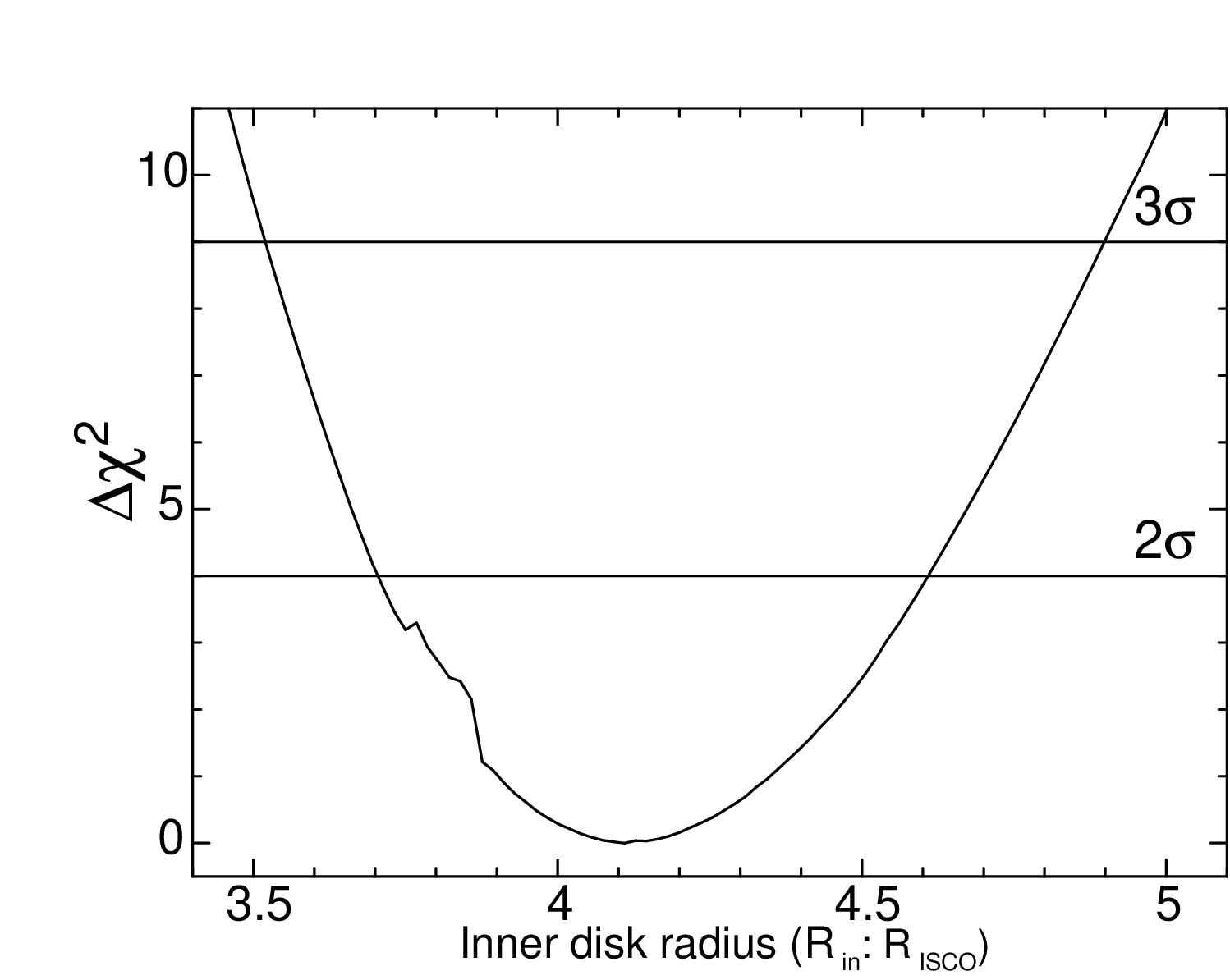}
\includegraphics[scale=0.22,angle=-90]{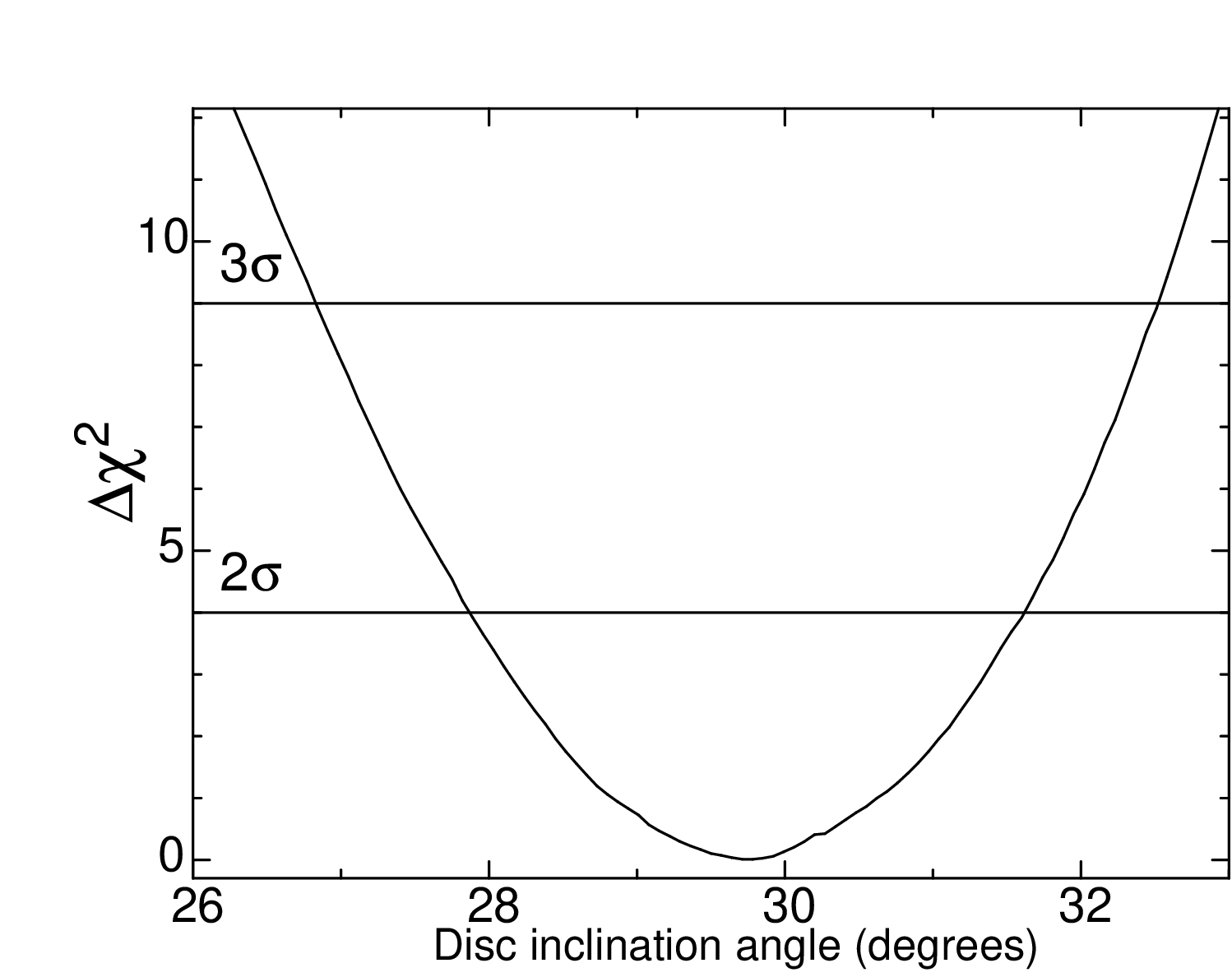}
\includegraphics[scale=0.22,angle=-90]{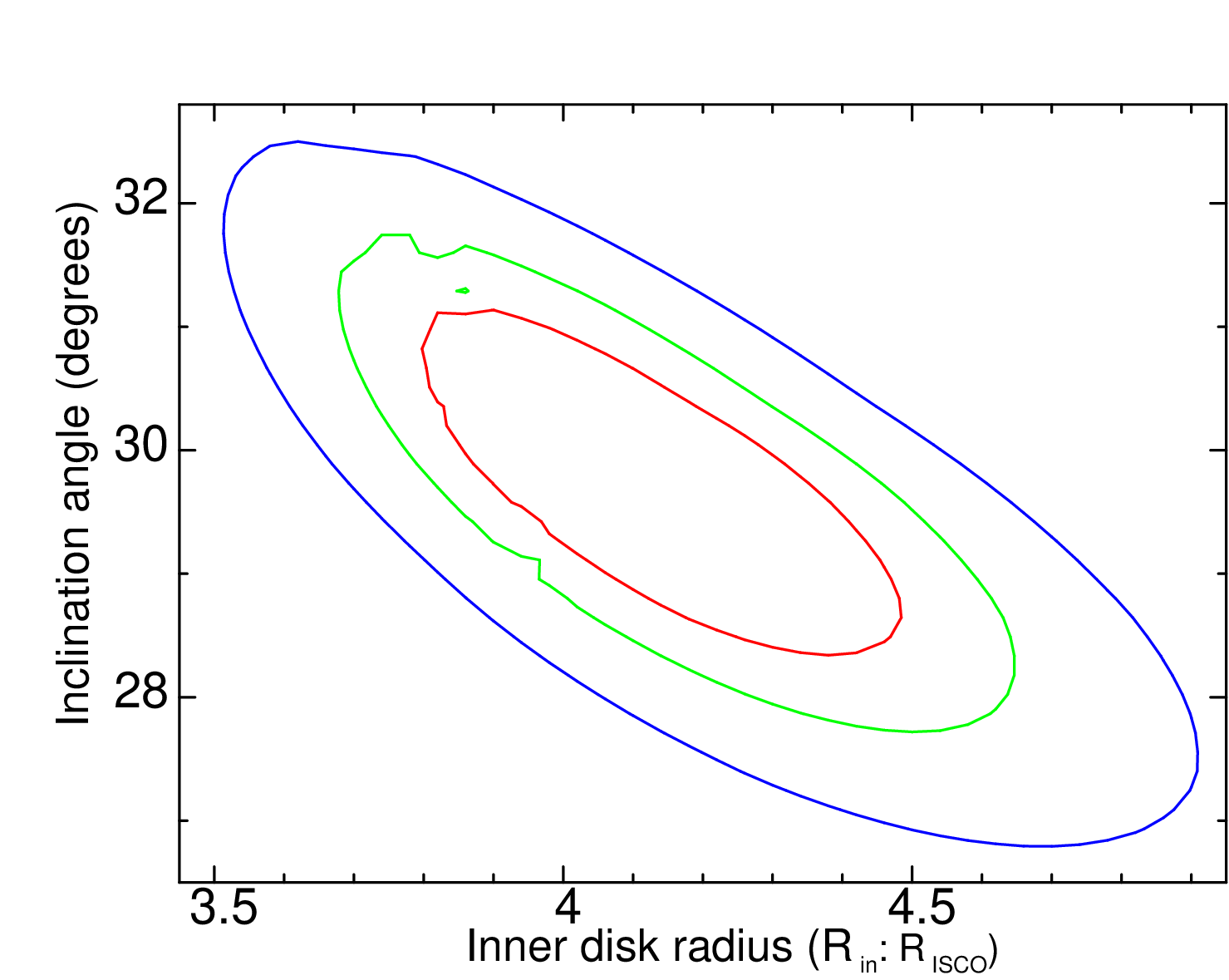}
\caption{The plot shows the constraints on inner disk radius R$_{in}$ and inclination angle {\it i}. The values of R$_{in}$ and {\it i} are given by the best fit model {\tt TBabs$\times$(diskbb + relxill(lp)Cp)}. Using best fit model, correlation between R$_{in}$ and {\it i} is explored. In the right panel 3$\sigma$, 2$\sigma$ and 1$\sigma$ confidence contours are shown in blue, green and red, respectively.}
\label{fig:fig5}
\end{figure*}

\hspace{0.8cm} After modeling the broadband spectrum, the non-thermal power-law index is found to be 1.51$^{+0.01}_{-0.01}$, which indicates that MAXI J1820+070 is in the hard state. During the hard state, a variable power-law continuum is often detected in black hole X-ray binaries. The non-thermal photons contributing to the variable power-law continuum originates from a spatially compact region called the ``corona". The exact geometry of the corona is still not very clear, but it is believed that the corona is located above the central region of the accretion disk \citep{Reynolds2014} and powered by the magnetic field associated with the accretion disk \citep{Galeev1979}. The hot power-law source irradiated the accretion disk, which produces fluorescent and backscattered radiation from the accretion disk. The re-emission from the irradiated disk produces the reflection spectrum \citep{Fabian2010, Fabian2016}.

We found a lamp-post height (h) of 3.9$^{+0.4}_{-0.3}$ r$_{g}$ and an electron temperature (kT$_{e}$) of 118$^{+34}_{-27}$ keV, suggesting a compact and relatively hot corona, which is the source of hot electrons required for Compton scattering and eventually for reflection characteristics.

The reflection spectrum normally includes a soft excess $\sim1$ keV, broad iron line, and Compton hump around 30 keV \citep{Fabian2016}. The soft excess around 1 keV is due to disk thermal emission \citep{Ross2005}.It is believed that these features originate due to photoelectric absorption by the free and bound electrons in the accretion disk and Compton up-scattering of thermal disk seed photons with the hot electrons from the corona. Above 20 keV, Compton scattering dominates resulting in a Compton hump around 30 keV. Below 20 keV, photoelectric absorption becomes important and create a drop in the spectra \citep{Lightman1988}. In our study, we observed that the Fe-K$\alpha$ line is symmetrically broadened and single peaked. The single peak of the Iron line characterizes the inner disk emission \citep{Miller2009}. The detected Iron line is elongated from 6 to 7 keV. The Doppler and the general relativity (GR) effects along with the scattering of the photons from the hot inner flow are believed to be the cause of extended Fe-K$\alpha$ line \citep{Fabian2000, Miller2007}.

From the reflection spectroscopy, we observed that the iron abundance (A$_{Fe}$) is 4.9$^{+0.4}_{-0.3}$, relative to the solar values. The overabundance of Fe has been detected in various X-ray binaries \citep[e.g.][]{Degenaar2017, Garcia2018, Tomsick2018}. According to \citet{Reynolds2012}, radiative levitation can magnify iron abundances and rapidly increase the A$_{Fe}$ value in X-ray binaries.

\section*{acknowledgements}
\hspace{0.8cm}This research has made use of the software and/or data obtained through the High Energy Astrophysics Science Archive Research Center (HEASARC) online service, provided by the NASA/Goddard Space Flight Center and the {\it SWIFT} data center. We are thankful to the referee for providing constructive comments, which greatly improved the quality of the paper. We also thank {\it NuSTAR} team for making {\it NuSTAR} data public. {\it MAXI} data are obtained from MAXI team, RIKEN, JAXA. We are also very grateful to the proposers of this observation, without whom this research couldn't have been pursued. The authors are also thankful to Dr Mayukh Pahari, University of Southampton, for constructive discussions for diagnostics and interpretations.

\bsp
\label{lastpage}

\end{NoHyper}
\end{document}